\documentclass[aps,prl,showpacs,twocolumn,superscriptaddress]{revtex4}

\usepackage{graphicx}
\usepackage{dcolumn}
\usepackage{bm}

\begin{document}

\title{High-visibility nonclassical interference between pure heralded single photons and weak coherent photons   \\}

\author{Rui-Bo Jin}
\email{jin@quantum.riec.tohoku.ac.jp}
\affiliation{Research
Institute of Electrical Communication, Tohoku University, Sendai
980-8577, Japan}
\author{Jun Zhang}
\affiliation{Research Institute of Electrical Communication, Tohoku University, Sendai 980-8577, Japan}
\author{Ryosuke Shimizu}
\affiliation{PRESTO, Japan Science and Technology Agency, Kawaguchi 332-0012, Japan}
\author{Nobuyuki Matsuda}
\affiliation{Research Institute of Electrical Communication, Tohoku University, Sendai 980-8577, Japan}
\author{Yasuyoshi Mitsumori}
\affiliation{Research Institute of Electrical Communication, Tohoku University, Sendai 980-8577, Japan}
\author{Hideo Kosaka}
\affiliation{Research Institute of Electrical Communication, Tohoku University, Sendai 980-8577, Japan}
\author{Keiichi Edamatsu}
\affiliation{Research Institute of Electrical Communication, Tohoku University, Sendai 980-8577, Japan}

\date{\today }

\begin{abstract}

We present an experiment of nonclassical interference between a
pure heralded single-photon state and a weak coherent state. Our
experiment is the first to demonstrate that spectrally pure single
photons can have high interference visibility, 89.4 $\pm$ 0.5\%,
with weak coherent photons. Our scheme lays the groundwork for
future experiments requiring quantum interference between photons
in nonclassical states and those in coherent states.

\end{abstract}


\maketitle

Nonclassical interference between independent photons (NIBIP)
plays a very important role in quantum information processing.
One kind of such NIBIP is the interference between photons from
different spontaneous parametric down-conversion (SPDC) sources,
which is vital to the preparation of the multi-photon entangled
state \cite{Lu2007} needed for implementing quantum networks
\cite{Gisin2007} and quantum computing algorithms
\cite{Lanyon2007}.
Another kind of NIBIP is the interference between single photons
from SPDC and weak coherent, i.e., local oscillator (LO) photons
from the laser source. This kind of interference is fundamental
for homodyne detection \cite{Lvovsky2009}, and is also the key to
quantum optical catalysis \cite{Lvovsky2002} and quantum circuits
\cite{Pittman2003b, Pittman2005}.

The first experiment of nonclassical interference between heralded
single photons from SPDC and LO was carried out by Rarity \emph{et
al} in 1997 \cite{Rarity1997,Rarity2005}.
Since LO photons have no phase correlation with SPDC photons,
i.e., signal and idler photons, the sources in the experiment can
be thought as independent sources.
However, in general, the signal and idler photons generated from
SPDC have correlated frequencies, and thus the heralded single
photons based on SPDC are not pure in terms of their
spectrotemporal modes. This lack of purity inevitably degrades the
indistinguishability between the signal (or idler) and LO photons,
resulting in low interference visibility.
Traditionally, bandpass filters
were employed to improve the
indistinguishability and interference visibility.
Spectral filtering is one way to improve the indistinguishability
between signal and LO photons, but this method has the drawback of
severely decreasing the count rate.
Recent advances in the preparation of a pure single-photon source
help solve this problem.
When a phase-matching condition is carefully engineered, a pure
heralded single-photon state can be generated in SPDC crystals
\cite{U'Ren2005, Wasylczyk2007,Mosley2008a,Mosley2008b} and
photonic crystal fibers \cite{Fulconis2007,Halder2009,Cohen2009,
Soller2010}.

By using SPDC with the group velocity matching condition in a
potassium-dihydrogen-phosphate (KDP) crystal \cite{Mosley2008a},
we prepared an intrinsically pure heralded single-photon state,
which interfered  with a weak coherent state in a three-photon
Hong-Ou-Mandel (HOM) interference \cite{Hong1987} without spectral
filtering.
Our experiment is the first to demonstrate that spectrally pure
heralded single photons can have high-visibility interference with
weak coherent photons without any spectral filtering.

The two-photon component of the final state of SPDC
can be expressed as
\begin{equation}\label{eq1}
 \left| {\psi _{si} } \right\rangle  = \int\limits_0^\infty  {\int\limits_0^\infty  {d\omega _s } d\omega _i } f(\omega _s ,\omega _i )\hat a_s^\dag  (\omega _s )\hat a_i^\dag  (\omega _i )\left| 0
 \right\rangle,
\end{equation}
where
$ f(\omega _s ,\omega _i )=\phi (\omega _s ,\omega _i)\alpha (\omega _s +\omega _i )$
 is the joint spectral distribution function \cite{Grice2001}.
$\phi (\omega _s ,\omega_i )$ and $\alpha (\omega _s +\omega _i )$
are the phase-matching function and the pump envelope function,
and the subscripts $s$ and $i$ denote signal and idler photons,
respectively.
By carefully choosing the phase-matching condition,
as described below,
the joint spectral distribution function of the signal and idler
photons can attain a factorable state \cite{Wasylczyk2007}, which
satisfies
\begin{equation}\label{eq2}
f(\omega _s ,\omega _i ) = g_s (\omega _s )  g_i (\omega _i).
\end{equation}

\begin{figure}[tbp]
\includegraphics[width= 0.45 \textwidth]{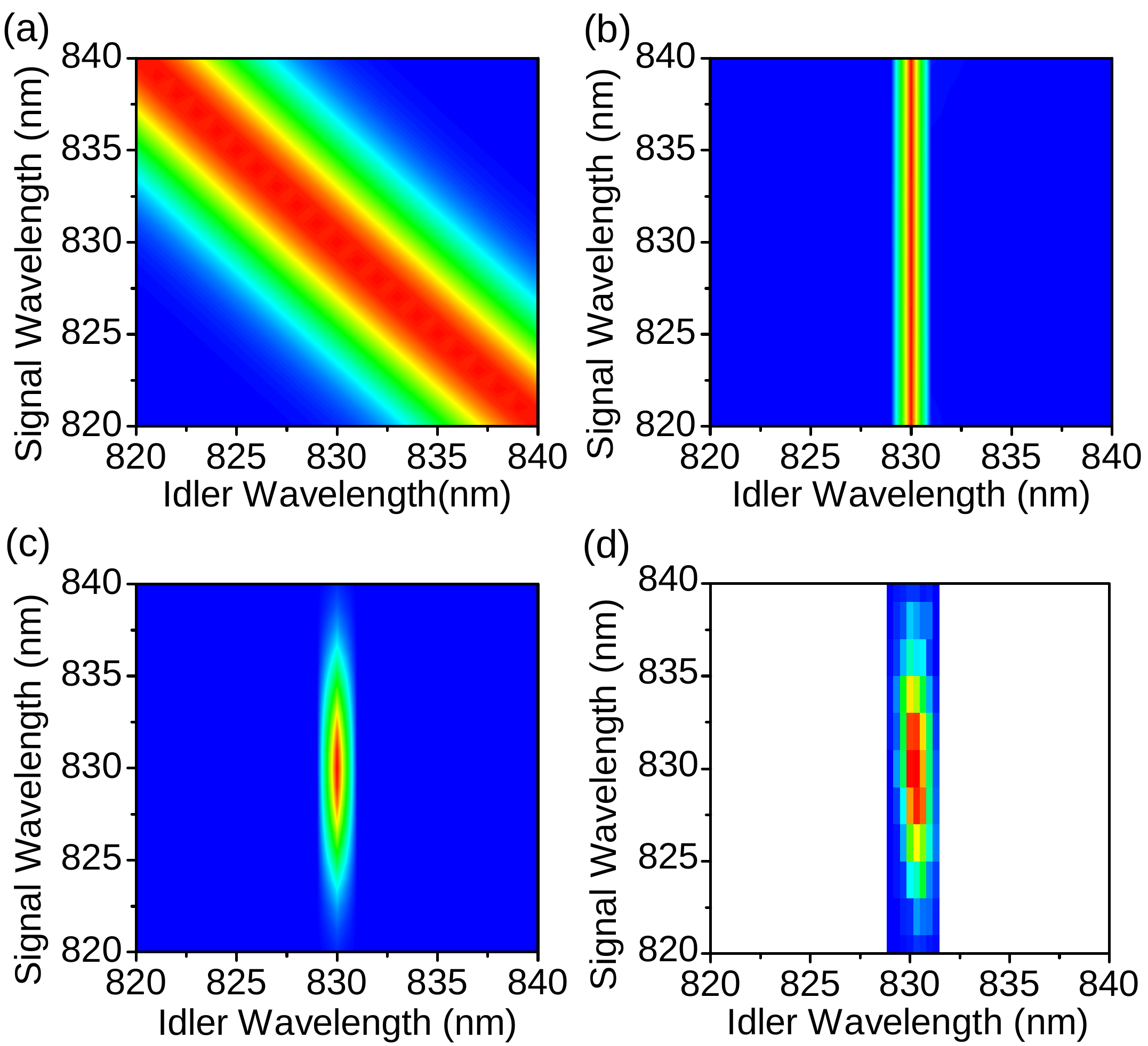}
\caption{ Density plots of the (a) pump envelope function, (b)
phase-matching function, (c) calculated joint spectral
distribution function, and (d) experimentally observed joint
spectral distribution, of the SPDC we employed. } \label{JSD}
\end{figure}

The purity of the signal is defined as
$\gamma\equiv {\rm Tr}(\hat\rho_s^2)$,
where
$\hat \rho_s={\rm Tr}_i(\left| {\psi _{si}
}\right\rangle\left\langle{\psi _{si} }\right| )$ is the reduced
density operator of the signal.
This purity is determined by the factorability of the joint
spectral distribution $f(\omega _s ,\omega_i )$ \cite{Mosley2008a}
and can be calculated numerically using Schmidt decomposition
\cite{U'Ren2005}.
In the case of the KDP crystal, the group velocity (GV) of the
415-nm pump (e-ray) equals the GV of the 830-nm signal (o-ray),
and is far from the GV of the 830-nm idler (e-ray).
Under this condition, the signal and idler are in a factorable
state \cite{ Wasylczyk2007}. Figures\,\ref{JSD} (a-c) present the
theoretical calculation of the (a) pump envelope function $|\alpha
(\omega _s +\omega _i )|^2$, (b) phase-matching function $|\phi
(\omega _s ,\omega_i )|^2$, and (c) joint spectral distribution
$|f(\omega _s ,\omega_i )|^2$, respectively, when a pump beam (415
nm, FWHM = 2.3 nm) is focused on a 15-mm-long KDP crystal.
It is obvious that (a) is frequency entangled; however, (b) is
sharp and functions as a delta function. As a result, the product
of (a) and (b) is factorable, as shown in (c).
Fig.\,\ref{JSD} (d) is the experimentally measured joint spectral
distribution, and will be explained in detail later.

\begin{figure}[tbp]
\includegraphics[width = 0.25 \textwidth]{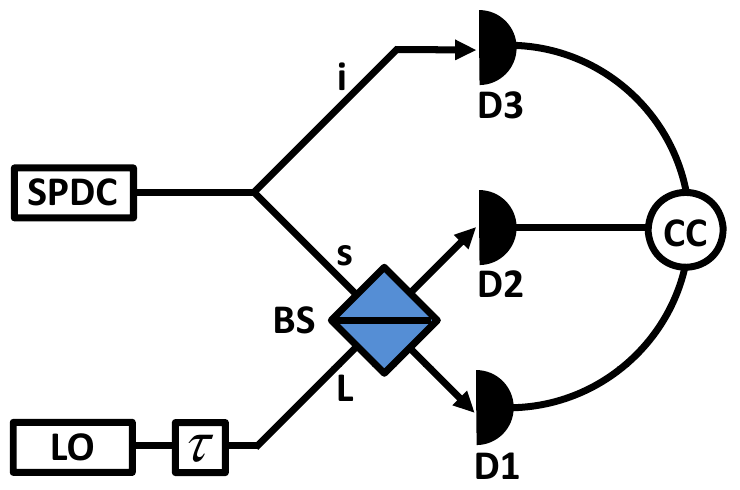}
\caption{ Schematic model of the experiment. The LO photon (L),
after a delay $\tau$, interfered with the signal (s) at the beam
splitter (BS) with the idler (i) as a heralder. These photons were
detected by three detectors and recorded by a three-fold
coincidence counter (CC). } \label{3fold}
\end{figure}

Next we considered the interference between the signal and LO
photons, with the idler as the heralder, as shown in
Fig.\,\ref{3fold}.
If both the signal and LO were single photons that are
indistinguishable from each other, we might expect a normal HOM
interference \cite{Hong1987}.
However, in our case, the signal could be treated as a single
photon when heralded by the sister idler photon.
Thus, three-fold coincidence is necessary to ensure that a single
signal photon interferes with an LO photon.
In addition, the mean photon number in an LO pulse should be low
enough that the probability of finding more than two LO photons is
negligible.
Another essential factor in this experiment is the
indistinguishability between the signal and LO photons.
Not only the identity in spectrotemporal profiles but also their
purity, as described above, are essential to ensure
indistinguishability \cite{Laiho2009,Cassemiro2010}.

\begin{figure}[tbp]
\includegraphics[width = 0.25 \textwidth]{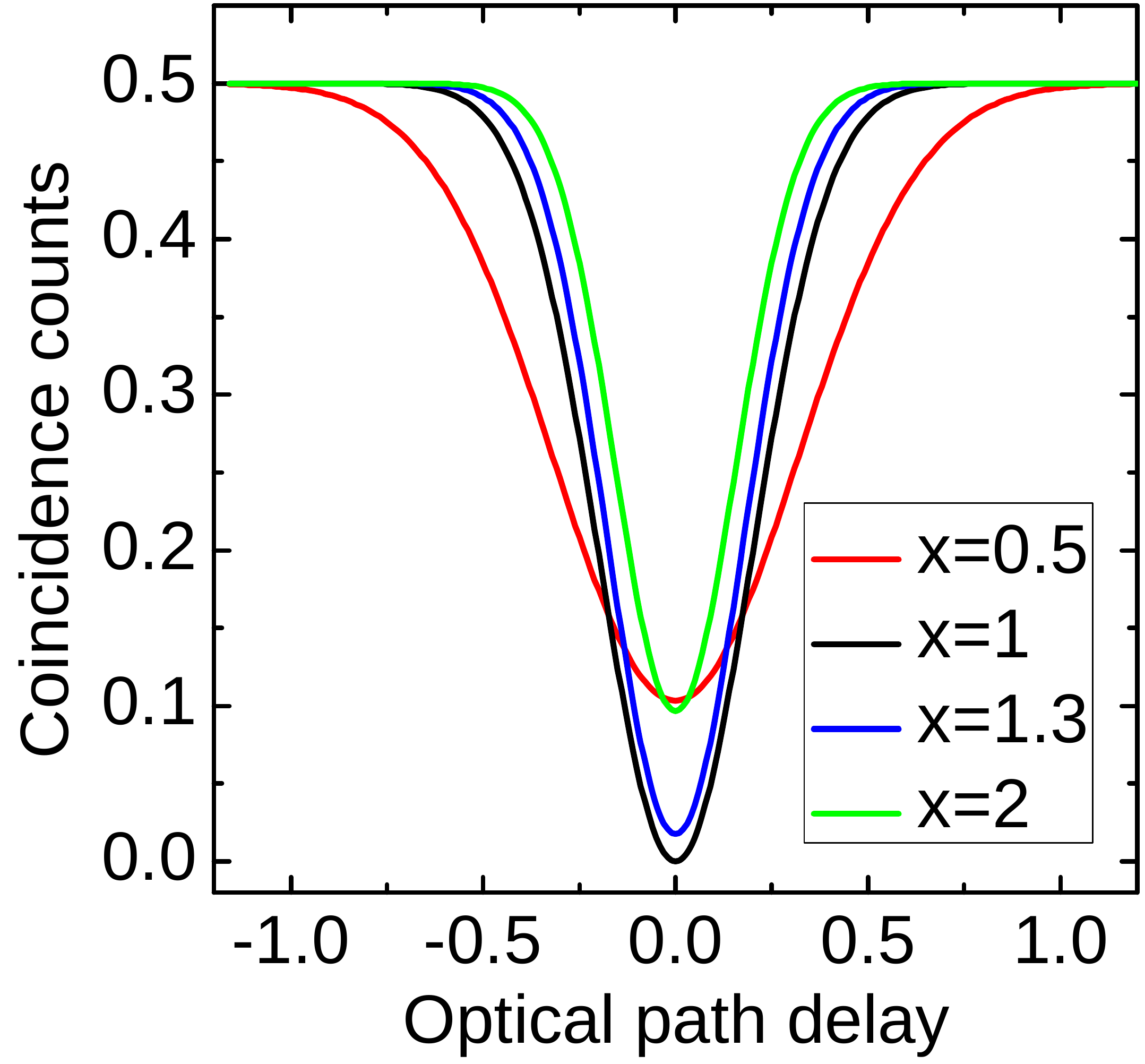}
\caption{Theoretical calculation of HOM interference between pure
signal and LO photons with identical central wavelengths
($\delta$=0) and different bandwidth ratios
($x$=$\sigma_s/\sigma_L$). For $x=1$,  we obtain $V=1$. When
$x=1.3$, 2 and 0.5,   $V=0.96$, 0.81, 0.79, respectively. }
\label{tdip}
\end{figure}

Assuming that the spectrotemporal modes of both the signal and LO
are pure, the three-fold coincidence count between the signal,
idler and LO as a function of the delay $\tau$ between the signal
and LO can be expressed as \cite{Ozdemir2002,Mosley2007}
\begin{equation}\label{equ}
P(\tau ) = \frac{1}{2} - \frac{{\sigma _s \sigma _L }}{{\sigma
_s^2  + \sigma _L ^2 }} \exp \left[ - \frac{{\sigma _s ^2 \sigma
_L ^2\tau ^2  + 4\delta^2 }} {{2(\sigma _s ^2 + \sigma _L ^2 )}}
\right],
\end{equation}
where $\sigma _s$ and  $\sigma_L$ are the bandwidths of Gaussian
spectra for the signal and LO, respectively, and $\delta$ is the
central frequency difference between the signal and LO.
When $\delta$=0, the interference visibility $V$ is written as
\begin{equation}\label{c}
V\equiv \frac{P(\infty)-P(0)}{P(\infty)}=\frac{2x}{ 1+ x^2}={\rm sech}\, \xi,
\end{equation}
where $x=\sigma_s/\sigma_L$ and $\xi=\ln(x)$. Perfect
interference, or $V$=1, is obtained, when $\delta$=0 and $x$=1.
Figure \ref{tdip} shows the calculated HOM interference pattern
$P(\tau$) for $\delta$=0 and some different values of $x$. We note
that $V$ is still as large as 0.96 when $x$=1.3, indicating that a
small difference between the bandwidths of the LO and signal
photons does not have a large influence on the interference
visibility.

\begin{figure}[tbp]
\includegraphics[width= 0.45 \textwidth]{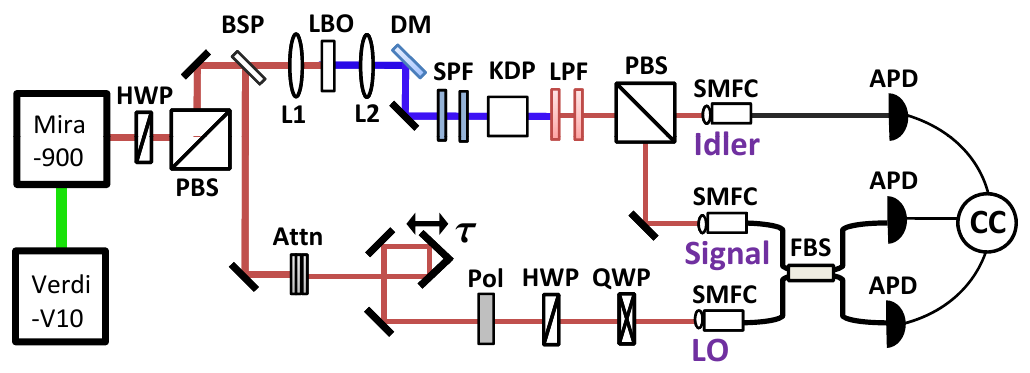}
\caption{ The experiment setup.  CC=coincidence counter,
APD=avalanche photodiodes, FBS=fiber beam splitter, SMFC=single
mode fiber coupler, PBS=polarizing beam splitter, QWP=quarter wave
plate, HWP=half wave plate, Pol=polarizer, Attn=attenuator,
BSP=beam sampler, DM=dichroic mirror, SPF=short wave pass filter,
LPF=long wave pass filter.} \label{setup}
\end{figure}

The experimental setup is displayed in Fig.\,\ref{setup}.
Femtosecond pulses (temporal duration$\sim$ 150 fs, center
wavelength=830 nm, FWHM = 7.1 nm) from the mode-locked Titanium
sapphire laser (Coherent, Mira900) were frequency-doubled by an
0.8-mm-thick lithium triborate (LBO) crystal and were used as the
pump source for the SPDC.
Pump pulses with power of 60 mW passed through a 15-mm-long KDP
crystal cut for type-II (eoe) degenerate phase-matching at 830 nm
($\theta$ = 67.8$^\circ$ ).
The down-converted photons, i.e., the signal (o-ray, FWHM = 9.3
nm) and idler (e-ray, FWHM = 1.9 nm) were separated by a
polarizing beam splitter.
Then, idler photons were coupled into a single-mode fiber, and
signal photons were coupled into a 50:50 single-mode fiber beam
splitter (FBS) (Thorlabs, FC830-50B-FC).
Fundamental laser pulses reflected from a beam sampler and highly
attenuated by neutral density filters were used as LO photons.
The polarization of the LO was adjusted by a polarizer, a
half-wave plate, and a quarter-wave plate so that we could obtain
the highest possible interference visibility between the signal
and LO.
Finally, all the collected photons were sent to three silicon
avalanche photodiode (APD) detectors (PerkinElmer, SPCM-AQRH14)
connected to a three-fold coincidence counter.

To check the factorability and purity of the prepared SPDC photon
pairs, we measured the joint spectral distribution by putting a
pair of monochromators on the signal and idler arms.
The coincidence counts between signal and idler were recorded
while scanning the wavelengths of the two monochromators.
The measured joint spectral distribution is shown in
Fig.\,\ref{JSD} (d).
The Schmidt value \cite{U'Ren2005, Mosley2008a} calculated from
Fig.\,\ref{JSD} (d) was  $1.03$, showing good factorability
\cite{Mosley2008b}, which ensures the high purity of the state we
prepared.

\begin{figure}[tbp]
\includegraphics[width = 0.45 \textwidth]{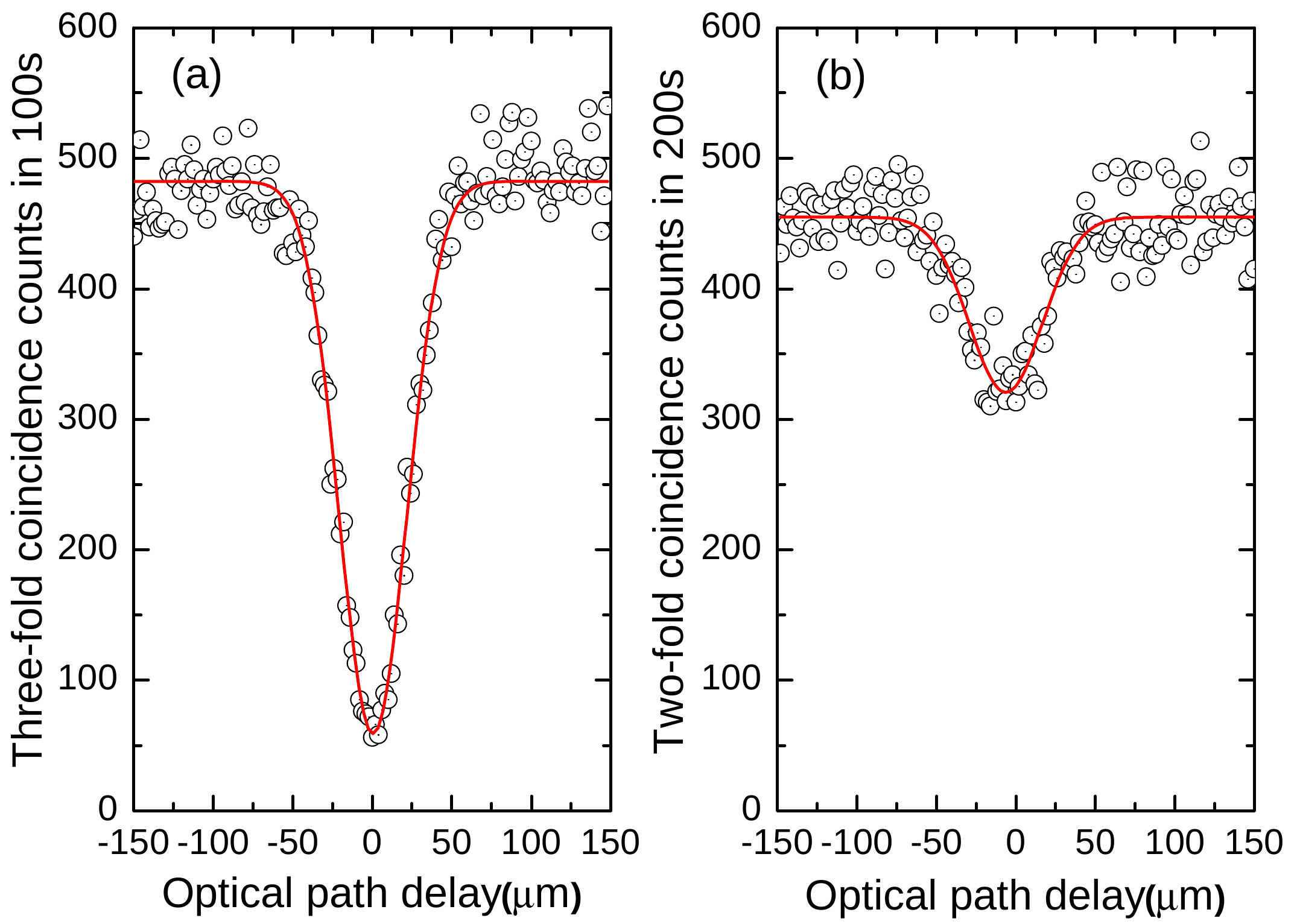}
\caption{ Observed HOM interference. (a) Three-fold coincidence
counts between the LO and heralded signal, with the idler as the
heralder. (b) Two-fold coincidence counts between the LO and
signal, without the heralder. No background signals were
subtracted in either (a) or (b). The solid curves represent
Gaussian fits to the data points. } \label{2dip}
\end{figure}

Figure\,\ref{2dip} (a) shows the result of the three-fold
coincidence count rate as a function of the optical path delay
$\tau$.
The observed single count rates of the idler, signal and LO were 9
kHz, 9 kHz and 600 kHz, respectively.
In this case, the average photon number per LO pulse was less than
0.02.
The two-fold coincidence count rate between the signal and idler
was 1.2 kHz, while the three-fold coincidence count rate between
the  signal, idler and LO was 4.8 Hz.
The three-fold counting rate exhibited a steep HOM dip around $\tau=0$,
as predicted in Fig.\,\ref{tdip}.
The maximum visibility observed was 89.4 $\pm$ 0.5 \%, with an
FWHM of 50.1 $\mu$m.
In contrast, as shown in Fig.\,\ref{2dip} (b), the visibility of a
two-fold coincidence count between the signal and LO was only 29.5
$\pm$ 0.3\%.
In this measurement, the corresponding single count of both the
signal and LO were 12 kHz.

With the idler as a heralder, the interference between the signal
and weak LO can be viewed as an interference between two
single-photon states, which can achieve 100\% visibility in the
ideal case.
On the contrary, without the heralding by the idler, the two-fold
interference is only a classical interference between a thermal
signal state and a weak coherent LO state, and  the upper limit of
the visibility is only 50\%.
This is the reason for the much higher visibility in
Fig.\,\ref{2dip} (a) than in Fig.\,\ref{2dip} (b).

In the experiment, the FWHM of the signal and LO spectra were 9.3
nm and 7.1 nm, respectively. According to Eq.\,(\ref{equ}),
the FWHM of the three-fold HOM dip was expected to be 44.5
$\mu$m, which is in reasonable agreement with the experimental
value of 50.1 $\mu$m.
The slightly longer value in the experiment might originate from
stretched UV pump duration caused by group velocity dispersion
(GVD) in the SHG crystal.
We also expect,
using Eq.\,(\ref{c}) and the measured FWHMs,
that $V$=96.5 \%.
The measured visibility  89.4 $\pm$ 0.5 \% was slightly smaller
than the theoretically expected value. 
The result may derive from the GVD effect in the SHG crystal
\cite{Rarity1997}, and the background accidental counts.
Nevertheless, in comparison with the first experiment of
nonclassical interference from independent sources
\cite{Rarity1997}, which employed a 3-nm bandpass filter and
achieved a visibility of 62.8 $\pm$ 1.2 \% in three-fold and 4.6
$\pm$ 0.2 \% in two-fold HOM interferences, we achieved
significant improvement not only in the visibility but also in the
efficiency, using the spectrally pure single-photon source.

Mosley \emph{et al.} \cite{Mosley2008a, Mosley2008b} demonstrated,
for the first time, that pure heralded single photons can be
generated through group velocity-matched SPDC, without spectral
filtering.
In their experiment, two independent KDP crystals were
pumped to produce two identical pairs of photons.
 The observed interference visibilities were 94.4 $\pm$ 1.6 \% between idlers
and 89.9 $\pm$ 3.0 \% between signals.
It should be emphasized that our scheme was different from their
experiment.
Both interfering photons in Refs.\,\cite{Mosley2008a, Mosley2008b}
were in heralded single-photon states, while in our approach, one
source was in a heralded single-photon state, and the other was in
a weak coherent state.
Our experiment manifested that spectrally pure single-photon
states can exhibit high-visibility nonclassical interference even
with classical, weak coherent states.

Many subareas of quantum information processing
\cite{Lu2008,Kaltenbaek2006, Zavatta2006, Rarity2003, Lanyon2007,
Pittman2005, Pittman2004, Pittman2001, Pittman2003b, Pittman2003a}
require nonclassical interference of photons from independent
sources.
Traditionally in these experiments, spectral filtering has been
utilized to improve visibility at the expense of decreasing event
efficiency.
When the system expands to utilize more photons, this may become a
severe problem.
With the scheme proposed in this paper, we can improve the
visibility without such expense, so that  the system has a better
expandability.

Another application of our approach is homodyne-based quantum
metrology and quantum information protocols.
Homodyne detection is a widely used technique in quantum optics,
in which a quantum signal mixes with a strong LO on a beam
splitter.
Conventionally, the LO and signal are filtered by narrow bandpass
filters to match the modes and improve their indistinguishability
\cite{Lvovsky2009}.
The use of bandpass filters, of course, decreases the event
efficiency, increasing the duration of the acquisition process.
To date, preparing the signal in a pure spectrotemporal mode, and
at the same time matching the modes of the LO and signal is still
a challenging task \cite{Lvovsky2009}.
The high-visibility interference between the signal and LO in our
experiment provides a good solution to the mode-matching problem
in homodyne detection.
In addition, the recent proposal on the preparation of high-NOON
states by mixing SPDC photons with coherent photons
\cite{Afek2010} also highlights the need for a pure SPDC source
for spectrotemporal mode matching.

In conclusion,
we have experimentally demonstrated high-visibility nonclassical
interference between a spectrally pure heralded single-photon
state and a weak coherent LO state.
The observed three-fold HOM interference exhibited a visibility of
89.4 $\pm$ 0.5 \%, which is superior to previous results, without
any spectral filtering.
Our scheme has promising applications in quantum metrology and
quantum information experiments requiring indistinguishability and
quantum interference between photons in nonclassical states and
those in coherent states.

R.-B. Jin is grateful to P. J. Mosley for helpful discussions.
This work was supported by a Grant-in-Aid for Creative Scientific
Research (17GS1204) from the Japan Society for the Promotion of
Science.


\end{document}